\input{aipcheck}

\documentclass[
    ,final            
  ]
  {aipproc}

\layoutstyle{6x9}

\usepackage{epsfig}

\begin{document}

\title{E$_6$SSM\footnote{Based on the talks presented by S.King and R.Nevzorov
at CICHEP II and ICHEP'06 respectively}}

\classification{12.60.Jv, 12.60.Cn, 12.60.Fr}
\keywords      {Supersymmetry; Extra gauge groups; Electroweak symmetry breaking.}

\author{S. F. King}{
  address={School of Physics and Astronomy, University of Southampton,\\
Southampton, SO17 1BJ, UK}
}

\author{S. Moretti}{
  address={School of Physics and Astronomy, University of Southampton,\\
Southampton, SO17 1BJ, UK}
}

\author{R. Nevzorov}{
  address={School of Physics and Astronomy, University of Southampton,\\
Southampton, SO17 1BJ, UK}
}

\begin{abstract}
In this talk we discuss an $E_6$ inspired supersymmetric (SUSY) model 
with an extra $U(1)_{N}$ gauge symmetry under which right-handed neutrinos 
have zero charge. In this exceptional supersymmetric standard model 
(E$_6$SSM) the $\mu$--term is generated dynamically after the electroweak
symmetry breaking. We specify the particle content of the model 
and argue that the presence of a $Z'$ and exotic particles predicted 
by E$_6$SSM allows the lightest Higgs boson to be significantly heavier 
than in the MSSM and NMSSM. Other possible manifestations of E$_6$SSM 
at the LHC are also discussed.
\end{abstract}

\maketitle


\section{Introduction}

The cancellation of quadratic divergences in the supersymmetric 
models does not allow to solve the hierarchy problem of the 
standard model (SM) entirely. Indeed, the superpotential of the
simplest supersymmetric extension of the SM --- minimal 
supersymmetric (SUSY) standard model (MSSM) contains a bilinear 
term $\mu H_d H_u$. In order to get the correct pattern of electroweak 
(EW) symmetry breaking the parameter $\mu$ is required to be of 
the order of electroweak or SUSY breaking scale. At the same time the 
incorporation of the MSSM into supergravity or Grand Unified 
theories (GUT) results in $\mu\sim M_X - M_{Pl}$, where $M_X$ and
$M_{Pl}$ are GUT or Planck scales respectively. This is the so--called
$\mu$--problem.

An elegant solution to the $\mu$ problem naturally arises in the framework 
of superstring inspired $E_6$ models. At the string scale 
$E_6$ can be broken directly to the rank-6 subgroup 
$SU(3)_C\times SU(2)_L\times U(1)_Y\times U(1)_{\psi}\times
U(1)_{\chi}$.
Two anomaly-free $U(1)_{\psi}$ 
and $U(1)_{\chi}$ symmetries of the rank-6 model are defined by:
$E_6\to SO(10)\times U(1)_{\psi},~SO(10)\to SU(5)\times U(1)_{\chi}$.
Near the string scale the rank-6 model can be reduced further to an effective 
rank--5 model with only one extra $U(1)'$ gauge symmetry. 
The extra $U(1)'$ gauge symmetry 
forbids an elementary $\mu$ term but allows interaction $\lambda S H_d H_u$
in the superpotential. 
The scalar component of the SM singlet superfield $S$ acquires a non-zero 
vacuum expectation value (VEV) breaking $U(1)'$ and giving rise
to an effective $\mu$ term.
Here we review a particular $E_6$ inspired supersymmetric model 
with an extra $U(1)_{N}$ gauge symmetry in which right handed neutrinos 
do not participate in the gauge interactions \cite{3}.

At collider energies the gauge group is:
\begin{equation}
SU(3)_C\times SU(2)_L\times U(1)_Y \times U(1)_N
\label{low}
\end{equation}
where the Standard Model is augmented by an additional 
$U(1)_N$ gauge group
which is defined so that right-handed neutrinos are neutral
under it and can be superheavy. This gauge group is supposed to descend from
an $E_6$ GUT gauge group which is broken at the GUT scale.
The $U(1)_N$ gauge group (a particular case of $U(1)'$)
is broken near the TeV energy scale giving rise
to a massive $Z'$ gauge boson which can be discovered at the LHC.

To ensure anomaly cancellation the low
energy particle content of the E$_6$SSM 
must include complete fundamental $27$ representations of $E_6$. 
Thus in addition to the three families of SM quarks
and leptons we predict
three families of exotic quark 
states $D_i,\overline{D}_i$ which 
carry a $B-L$ charge $\left(\pm\frac{2}{3}\right)$,
and singlet fields
$S_i$ which carry non-zero $U(1)_{N}$ charges and therefore survive
down to the EW scale. We also predict three families of states
$H_{1i}$ and $H_{2i}$ which have the quantum numbers of
Higgs doublets. 
We also require a further pair $H'$ and $\overline{H}'$ 
from incomplete extra $27'$ and $\overline{27'}$ representations 
to survive to low energies
in order to ensure gauge coupling unification.
Thus in addition to a $Z'$ the E$_6$SSM
involves extra matter beyond the MSSM.

\section{The Superpotential and Parameter Counting}

The superpotential of the E$_6$SSM involves a lot of new 
Yukawa couplings in comparison to the SM. In general these new 
interactions violate baryon number conservation and induce 
non-diagonal flavour transitions. To suppress baryon number 
violating and flavour changing processes one can postulate 
a $Z^{H}_2$ symmetry under which all superfields except 
one pair of $H_{1i}$ and $H_{2i}$ (say $H_d\equiv H_{13}$ 
and $H_u\equiv H_{23}$) and one SM-type singlet field 
($S\equiv S_3$) are odd. The $Z^{H}_2$ symmetry reduces 
the structure of the Yukawa interactions to:
\begin{equation} 
\begin{array}{c} 
W_{\rm E_{6}SSM}\simeq \lambda_i S(H_{1i}H_{2i})+\kappa_i
S(D_i\overline{D}_i) +f_{\alpha\beta}S_{\alpha}(H_d H_{2\beta})+ \\[4mm]
\tilde{f}_{\alpha\beta}S_{\alpha}(H_{1\beta}H_u)+W_{MSSM}(\mu=0)\,,
\end{array}
\label{essm3}
\end{equation}
where $\alpha,\beta=1,2$ and $i=1,2,3$\,. In Eq.~(\ref{essm3}) we 
ignore $H'$ and $\overline{H}'$ for simplicity. 
Here we define $\lambda \equiv \lambda_3$. The $SU(2)$ doublets $H_u$ 
and $H_d$ play the role of Higgs fields generating the masses of 
quarks and leptons after electroweak symmetry breaking (EWSB). 
Therefore it is natural to assume 
that only $S$, $H_u$ and $H_d$ acquire non-zero VEVs. If $\lambda$ 
or $\kappa_i$ are large at the grand unification (GUT) scale $M_X$ 
they affect the evolution of the soft scalar mass $m_S^2$ of the 
singlet field $S$ rather strongly resulting in negative values of 
$m_S^2$ at low energies that trigger the breakdown of the $U(1)_{N}$ 
symmetry. To guarantee that only $H_u$, $H_d$ and $S$ acquire a VEV 
we impose a certain hierarchy between the couplings $H_{1i}$ and 
$H_{2i}$ to the SM-type singlet superfields $S_i$: 
$\lambda\gg \lambda_{1,2},\,f_{\alpha\beta}$ and $\tilde{f}_{\alpha\beta}$. 

The masses of the fermion components of $H'$ and $\overline{H}'$ are 
induced by the term $\mu'H'\overline{H}'$ in the superpotential. 
The corresponding mass term is not involved in the process of EWSB. 
Therefore parameter $\mu'$ remains arbitrary. Gauge coupling 
unification requires $\mu'$ to be within $100\,\mbox{TeV}$. 
The masses of scalar components of $H'$ and $\overline{H}'$ are 
determined by the soft masses $m^2_{H}$ and $m_{\overline{H}'}^2$ 
as well as by $\mu'$ and the corresponding bilinear scalar coupling 
$B'$ in the scalar potential. Because $\mu'$ and $B'$ can be complex 
the spectrum of survival components of $27'$ and $\overline{27'}$ 
is determined by six parameters.

The superpotential (\ref{essm3}) contains 14 new Yukawa couplings 
as compared to the MSSM with $\mu=0$. They are accompanied by 
14 trilinear scalar couplings in the SUSY scalar potential. 
In addition the scalar potential of E$_6$SSM includes 13 soft SUSY 
masses: six masses of exotic squarks $m_{\tilde{D}_i}$ and 
$m_{\tilde{\overline{D}}_i}$, four masses of non--Higgs fields 
$m_{\tilde{H}_{1,\alpha}}$ and $m_{\tilde{H}_{2,\alpha}}$ 
($\alpha=1,2$) and three masses of SM singlet scalar fields 
$m^2_{S_i}$. Because Yukawa and trilinear scalar couplings can be 
complex the $Z^{H}_2$--symmetric E$_6$SSM involves 75 new parameters 
in comparison to the MSSM with $\mu=0$ which determine masses and 
couplings of extra fields. Thirty of them are phases. Some of these 
phases can be eliminated by the appropriate redefinition of new 
superfields. 

Although $Z^{H}_2$ eliminates any problem related with baryon number
violation and non-diagonal flavour transitions it also forbids all 
Yukawa interactions that would allow the exotic quarks to decay. 
Since models with stable charged exotic particles are ruled out by 
different experiments \cite{Hemmick:1989ns} the $Z^{H}_2$ symmetry 
must be broken. But the breakdown of $Z^{H}_2$ should not 
give rise to the operators leading to rapid proton decay.
There are two ways to overcome this problem. 
The resulting Lagrangian has to be 
invariant either with respect to $Z_2^L$ symmetry, under which all 
superfields except lepton ones are even, or with respect to $Z_2^B$
discrete symmetry, which implies that exotic quark and lepton 
superfields are odd whereas the others remain even. 
The terms in the superpotential which 
permit exotic quarks to decay and are allowed by the $E_6$ symmetry 
can be written in the following alternative forms, depending
on which discrete symmetry is imposed.
If $Z_2^L$ is imposed then the following couplings are allowed:
\begin{equation}
W_1=g^Q_{ijk}D_{i} (Q_j Q_k)+g^{q}_{ijk}\overline{D}_i d^c_j u^c_k\,,
\label{essm31}
\end{equation}
which implies that exotic quarks are diquarks.
If $Z_2^B$ is imposed then the following couplings are allowed:
\begin{equation}
W_2=g^E_{ijk} e^c_i D_j u^c_k+g^D_{ijk} (Q_i L_j) \overline{D}_k\,.
\label{essm32}
\end{equation} 
which implies that exotic quarks are leptoquarks. 
We assume that the violation of the $Z^{H}_2$ 
symmetry in the E$_6$SSM is mainly caused  by the Yukawa couplings 
of the exotic particles to the quarks and leptons of the third generations.
This assumption 
results in three (six) extra Yukawa couplings if exotic quarks are diquarks
(leptoquarks). Thus together with the trilinear scalar couplings this
would increase the total number of independent parameters by 12 (24)
degrees of freedom.

\section{Phenomenological implications}

The E$_6$SSM Higgs sector includes two Higgs doublets $H_u$ and $H_d$ 
as well as a SM--like singlet field $S$. After the breakdown of the gauge 
symmetry two CP-odd and two charged Goldstone modes in the Higgs sector are
absorbed by the $Z$, $Z'$ and $W^{\pm}$ gauge bosons so that only six
physical degrees of freedom are left. They represent three CP-even (as
in the NMSSM), one CP-odd and two charged Higgs states (as in the
MSSM). 


\begin{figure}[b]
\centerline{\includegraphics[height=.3\textheight]{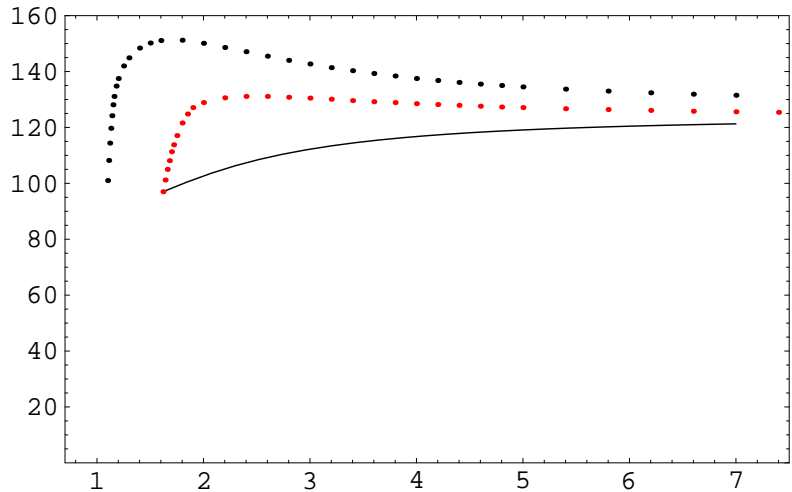}}\\
\caption{Two-loop upper bound on the lightest Higgs mass versus $\tan\beta$.
The solid, lower and upper dotted lines correspond to the theoretical restrictions
on the lightest Higgs mass in the MSSM, NMSSM and E$_6$SSM respectively.}
\label{essmfig1}
\end{figure}

As in any other SUSY model the mass of the lightest CP--even Higgs
boson $m_h$ in the E$_6$SSM is limited from above. In
Fig.~\ref{essmfig1} we plot the two-loop upper bounds on the mass of
the lightest Higgs particle in the MSSM, NMSSM and E$_6$SSM as a
function of $\tan\beta$. At moderate values of $\tan\beta$
($\tan\beta=1.6-3.5$) the upper limit on the lightest Higgs boson mass
in the E$_6$SSM is considerably higher than in the~ MSSM~ and~
NMSSM. It~ reaches~ the~ maximum~ value~ $150-155\,\mbox{GeV}$~ at~
$\tan\beta=1.5-2$ \cite{3}. 
The main reason for the increased Higgs mass in the E$_6$SSM
is due to the increased upper limit on the
coupling $\lambda$ (caused by the extra exotic states)
which controls the important F-term contribution to $m_h$.
At large $\tan\beta >10$ the
theoretical restriction on $m_{h}$ in the E$_6$SSM is
$4-5\,\mbox{GeV}$ larger than the one in the MSSM and NMSSM because of
the $U(1)_{N}$ $D$-term contribution to $m^2_{h}$. 
The discovery at
future colliders of a relatively heavy SM-like Higgs boson with mass
$140-155\,\mbox{GeV}$ will permit to distinguish the E$_6$SSM from the
MSSM and NMSSM.


\begin{figure}
\centerline{\epsfig{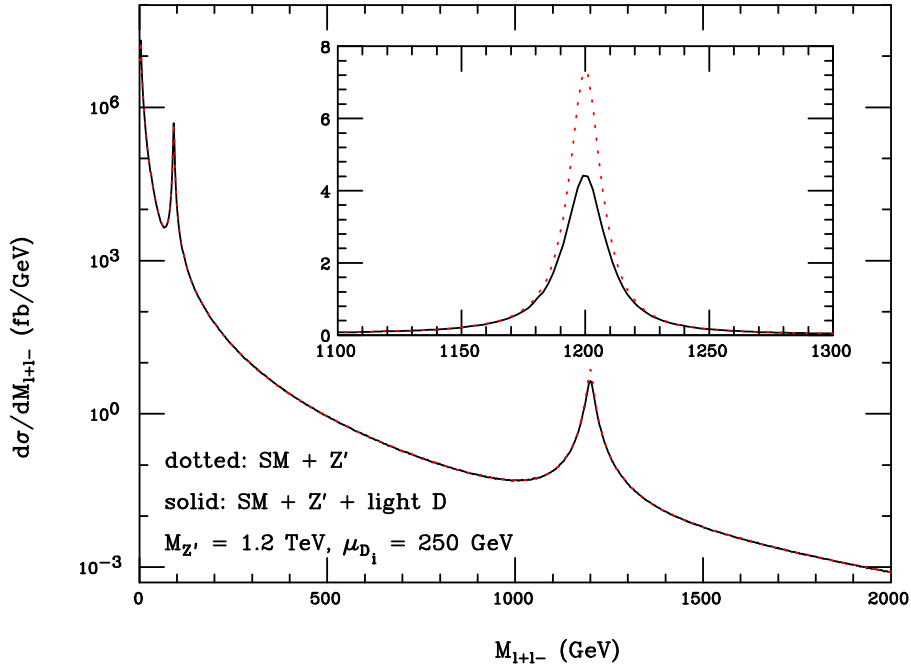}}
\caption{Differential cross section in the final state
invariant mass, denoted by $M_{l^+l^-}$, at the LHC for DY production
($l=e$ or $\mu$ only) in presence of a $Z'$ with and without the
(separate) contribution of exotic $D$-quarks with
$\mu_{Di}=250\,\mbox{GeV}$ for $M_{Z'}=1.2\,\mbox{TeV}$.}
\label{essmfig2}
\end{figure}


Other possible manifestations of our exceptional SUSY model at the LHC 
are related with the presence of a $Z'$ and of exotic multiplets of matter.
For instance, a relatively light $Z'$ will lead to enhanced production of 
$l^{+}l^{-}$ pairs ($l=e,\mu$). Fig.~\ref{essmfig2} shows the differential
distribution in invariant mass of the lepton pair $l^+l^-$ (for one
species of lepton $l=e,\mu$) in Drell--Yan production at the LHC with
and without light exotic quarks with representative masses of exotic quarks 
$\mu_{D_i}=250$ GeV for all three generations and with $M_{Z'}=1.2\,\mbox{TeV}$.
This distribution is promptly measurable at the CERN collider with a high
resolution and would enable one to not only confirm the existence of a
$Z'$ state but also to establish the possible presence of additional
exotic matter, by simply fitting to the data the width of the $Z'$
resonance. The analysis performed in \cite{Kang:2004bz} 
revealed that a $Z'$ boson in $E_6$ inspired models can be discovered at 
the LHC if its mass is less than $4-4.5\,\mbox{TeV}$. At the same time 
the determination of its couplings should be possible up to 
$M_{Z'}\sim 2-2.5\,\mbox{TeV}$ \cite{Dittmar:2003ir}.

\begin{figure}
\centerline{\epsfig{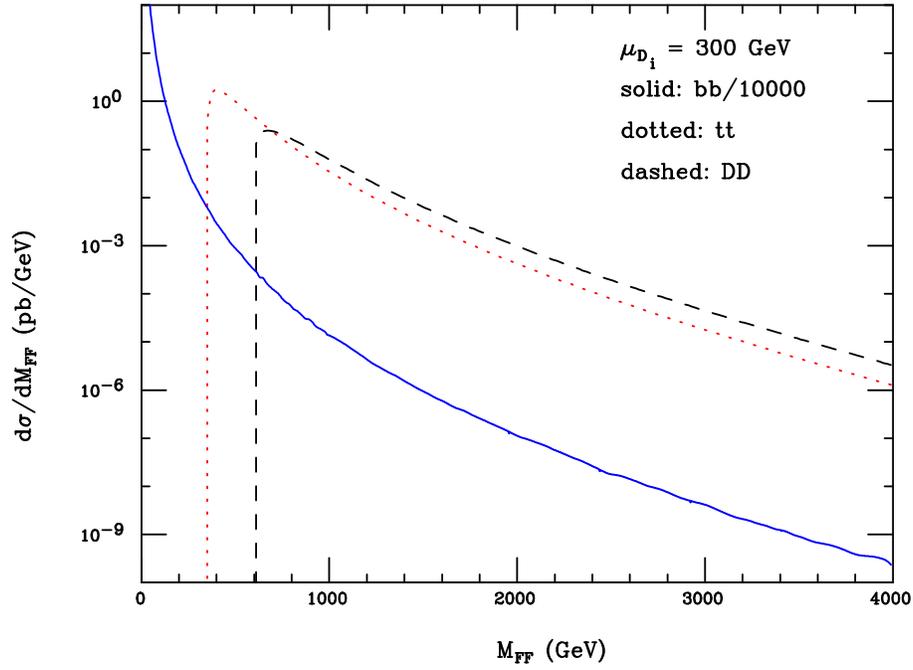}}
\caption{Cross section at the LHC for pair production of exotic $D$-quarks
as a function of the invariant mass of $D\overline{D}$ pair. Similar cross
sections of $t\overline{t}$ and $b\overline{b}$ production are also
included for comparison.}
\label{essmfig3}
\end{figure}

The exotic quarks can be also relatively light in the E$_6$SSM
since their masses are set by the Yukawa couplings $\kappa_i$ and 
$\lambda_i$ that may be small. Then the production cross section of 
exotic quark pairs at the LHC can be comparable with the cross section 
of $t\bar{t}$ production (see Fig.~\ref{essmfig3}). Since we 
have assumed that $Z_2^H$ is mainly broken by operators involving quarks and 
leptons of the third generation the lightest exotic quarks decay into 
either two heavy quarks $Q Q$ or a heavy quark and a lepton $Q\tau (\nu_{\tau})$, 
where $Q$ is either a $b$- or $t$-quark. This results in the growth of the 
cross section of either $pp\to Q\bar{Q}Q^{(')}\bar{Q}^{(')}+X$ or 
$pp\to Q\bar{Q}l^{+}l^{-}+X$. The discovery of the $Z'$ and exotic 
quarks predicted by the E$_6$SSM would represent a possible indirect 
signature of an underlying $E_6$ gauge structure at high energies 
and provide a window into string theory.

\begin{theacknowledgments}
The authors would like to to acknowledge support from the PPARC grant 
PPA/G/S/2003/00096, the NATO grant PST.CLG.980066 and the EU network 
MRTN 2004-503369.
\end{theacknowledgments}



\bibliographystyle{aipproc}   

\bibliography{sample}

\IfFileExists{\jobname.bbl}{}
 {\typeout{}
  \typeout{******************************************}
  \typeout{** Please run "bibtex \jobname" to optain}
  \typeout{** the bibliography and then re-run LaTeX}
  \typeout{** twice to fix the references!}
  \typeout{******************************************}
  \typeout{}
 }

\end{document}